\documentclass[twocolumn,showpacs]{revtex4}

\usepackage{graphicx}
\usepackage{dcolumn}
\usepackage{amsmath}

\makeatletter
\def\btt#1{\texttt{\@backslashchar#1}}%
\DeclareRobustCommand\bblash{\btt{\@backslashchar}}%
\makeatother

\begin{document}
\pagestyle{empty}
\preprint{UGe2 NMR}

\title{Evidence for Uniform Coexistence of Ferromagnetism and Unconventional Superconductivity in UGe$_2$: A $^{73}$Ge-NQR Study under Pressure}
\author{H. \textsc{Kotegawa}$^{1}$\thanks{Present address: Department of Physics, Faculty of Science, Okayama University, Okayama 700-8530, Japan}, A. \textsc{Harada}$^{1}$, S. \textsc{Kawasaki}$^{1}$, Y. \textsc{Kawasaki}$^{1}$\thanks{Present address: Department of Physics, Faculty of Engineering, Tokushima University, Tokushima 770-8506, Japan}, Y. \textsc{Kitaoka}$^{1}$, Y. \textsc{Haga}$^{2}$, E. \textsc{Yamamoto}$^{2}$, Y. \textsc{\=Onuki}$^{2,3}$, K. M. \textsc{Itoh}$^{4}$, E. E. \textsc{Haller}$^{5}$, and H. \textsc{Harima}$^{6}$}

\address{$^{1}$Department of Materials Engineering Science, Graduate School of Engineering Science, Osaka University, Toyonaka, Osaka 560-8531, Japan \\
$^{2}$Advanced Science Research Center, Japan Atomic Energy Research Institute, Tokai, Ibaraki 319-1195, Japan\\
$^{3}$Department of Physics, Graduate School of Science, Osaka University, Toyonaka, Osaka 560-0043, Japan \\
$^{4}$ Department of Applied Physics and Physico-Informatics,
Keio University, Yokohama 223-8522 Japan\\
$^{5}$ Department of Materials Science and Engineering
University of California at Berkeley and Lawrence 
Berkeley National Laboratory
Berkeley, CA 94720 USA\\
$^{6}$ Department of Physics, Faculty of Science, Kobe University, Nada, Kobe 657-8501, Japan}

\date{\today}

\begin{abstract}
We report on the itinerant ferromagnetic superconductor UGe$_2$ through $^{73}$Ge-NQR measurements under pressure ($P$). The $P$ dependence of the NQR spectrum signals a first-order transition from  the low-temperature ($T$) and  low-$P$ ferromagnetic phase (FM2) to high-T and high-P one (FM1) around a critical pressure of $P_x \sim 1.2$ GPa. The superconductivity exhibiting a maximum value of $T_{sc}=0.7$ K at $P_x \sim 1.2$ GPa, was found to take place in connection with the $P$-induced first-order transition. The nuclear spin-lattice relaxation rate $1/T_1$ has probed the ferromagnetic transition,~exhibiting a peak at the Curie temperature as well as a decrease without the coherence peak below $T_{sc}$. These results reveal the uniformly coexistent phase of ferromagnetism and unconventional superconductivity with a line-node gap. We remark on an intimate interplay between the onset of superconductivity and the underlying electronic state for the ferromagnetic phases.
\end{abstract}

\vspace*{5mm}

\maketitle

\section{Introduction}
The coexistence of magnetism and superconductivity (SC) is an important recent topic in condensed-matter physics. The uniformly coexistent phases of antiferromagnetism (AFM) and SC have been reported in U$M_{2}$Al$_{3}$ ($M={\rm Pd, Ni}$),\cite{Geibel} CeCu$_2$Si$_2$,\cite{Kitaoka01,YKawasaki01,YKawasaki02} CeIn$_{3}$,\cite{Mathur,SKawasaki01,SKawasaki03} and CeRhIn$_{5}$,\cite{Hegger,Mito01,SKawasaki02} where $f$-electrons are anticipated to contribute to both AFM and SC. 
~Recently, in a ferromagnet UGe$_2$ with a Curie temperature $T_{Curie}=52$ K at ambient pressure ($P=0$), $P$-induced SC was discovered to emerge under $P=1-1.6$ GPa \cite{Saxena,Huxley}. It is noteworthy that SC in UGe$_2$ disappears above $P_c\sim 1.6$ GPa beyond which ferromagnetism (FM) is suppressed. This fact implies that SC and FM in this compound may be cooperative phenomena.
~It is, however, currently believed that the uniformly coexistent phase of FM and SC is unlikely to exist because the Cooper pairs feel a non-vanishing internal field to prevent the onset of a spin-singlet SC.
It is,~therefore, surprising that both of FM and SC are carried by $5f$ electrons of uranium atoms and SC coexists with FM with a large moment of the order of 1$\mu_B$/U, which suggests that a spin-triplet pairing state may be formed.
Although SC and FM have been reported in HoMo$_{6}$S$_{8}$,\cite{HoMoS} ErRh$_{4}$B$_{4}$ \cite{ErRhB} and ErNi$_2$B$_2$C,\cite{ErNiBC} we should note that $T_{Curie} < T_{sc}$ and the two orders are competing in these cases.  FM is carried by localized $4f$ electrons of Ho and Er atoms, whereas SC by conduction electrons. 
In this context, the recent discoveries of SC in ferromagnets UGe$_2$,\cite{Saxena,Huxley} ZrZn$_2$, \cite{Pfleiderer} and URhGe \cite{Aoki} have been a great surprise.

\begin{figure}[h]
\centering
\includegraphics[width=8.5cm]{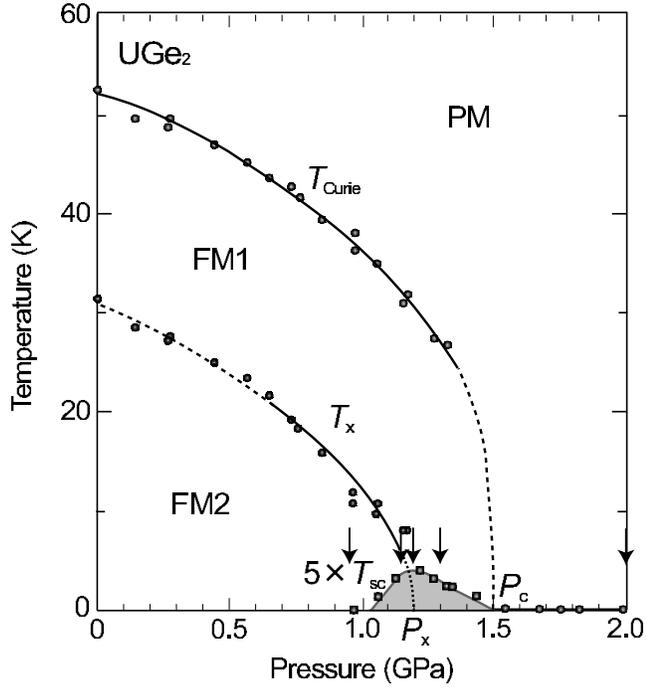}
\caption[]{The pressure versus temperature phase diagram of UGe$_2$ \cite{Kobayashi,Haga}. Arrows show values of $P$ where the present NQR measurements have been done. 
}
\end{figure}

Figure~1 shows the $P$ versus temperature ($T$) phase diagram of UGe$_2$ established from various measurements \cite{Saxena,Huxley,Kobayashi,Haga,Tateiwa,Pfleiderer2}. 
$T_{Curie}$ decreases monotonously from $T_{Curie}=52$ K at $P=0$ with increasing $P$.  SC sets in at pressures exceeding $P\sim 1.0$ GPa, exhibiting a maximum value of superconducting transition temperature $T_{sc}\sim 0.7-0.8$ K around $P_x\sim 1.2$ GPa.
The measurements of magnetization and resistivity show anomalous behaviors at $T_x$ far below $T_{Curie}$ \cite{Huxley,Tateiwa}. The high-$T$ phase is denoted as FM1 and the low-$T$ one as FM2 as indicated in the phase diagram.
An interesting point is that $T_{sc}$ becomes maximum at $P_x \sim 1.2$ GPa, where $T_x$ disappears as if it were a termination point of a first-order transition from FM2 to FM1 \cite{Pfleiderer2}. The fact that both SC and FM disappear simultaneously at $P_c\sim 1.6$ GPa suggests that SC is in a spin-triplet pairing state under the background of FM. However, it has been suspected from the measurements of diamagnetic susceptibility that both phases do not coexist but rather compete with each other; as $P$ increases, a volume fraction of SC grows over the whole system, whereas FM seems to become spatially inhomogeneous \cite{Motoyama}. This result raises a question as to whether or not the uniformly coexistent phase of SC and FM is realized in UGe$_2$.

In this article, we report on results of a series of nuclear-quadrupole-resonance (NQR) measurements of the enriched $^{73}$Ge that address the microscopic characteristics of SC and FM in UGe$_2$. We  have also examined to what type of superconducting order parameter exists in UGe$_2$.

\section{Experimental Procedures}

A polycrystalline sample enriched in $^{73}$Ge was prepared and crushed into powder for NQR measurements. Hydrostatic pressure was applied by utilizing a BeCu piston-cylinder cell, filled with Daphne oil (7373) as a pressure-transmitting medium. To calibrate a value of $P$ at a sample position at low temperatures, the $P$-dependence of $T_{sc}$ of Pb was measured by resistivity measurement. Furthermore, in order to inspect a pressure gradient in the cell, we have measured a $P$ increasing rate $d\nu_Q/dp$ of quadrupole frequency $\nu_Q$ of a reference sample (CeCu$_2$Si$_2$) with a narrowest linewidth ($\Delta\nu_Q\sim 0.01$ MHz) in NQR spectrum to date. In this sample, as $P$ increases, $\nu_Q$ increases linearly due to a linear increase of the electric field gradient at the Cu site that is caused by an increase in the lattice density. Using $d\nu_Q/dp = 9.52$ Hz/bar, a gradual increase of $\Delta \nu_Q$ with increasing $P$ assures that a possible distribution of $P$, $\Delta P/P$ is less than 3\% in the cell for NQR measurements. A $^{3}$He-$^{4}$He dilution refrigerator is used to reach the lowest temperature of $\sim 50$ mK. The NQR experiment was performed by the conventional spin-echo method under zero field in the frequency ($f$) range of 5 - 12 MHz. 

\section{Experimental Results}
\subsection{$^{73}$Ge-NQR spectrum $-$ Evidence for first-order transition$-$} 
\begin{figure}[h]
\centering
\includegraphics[width=8.5cm]{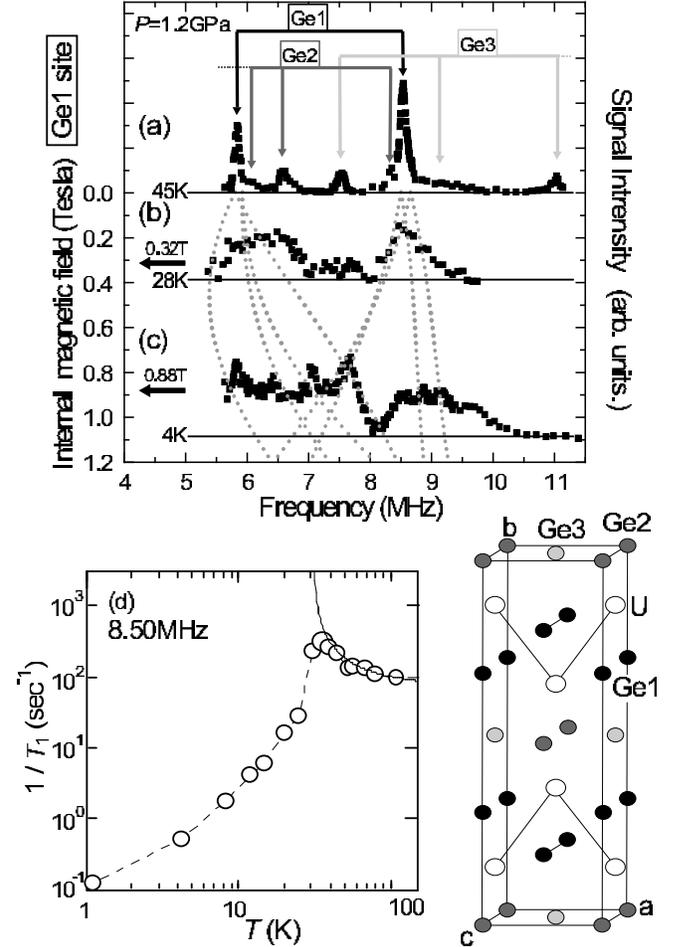}
\caption[]{The temperature dependence of NQR spectra at $P=1.2$ GPa. The NQR spectrum (a) at $T=45$ K for the paramagnetic (PM) state reveals well separated peaks associated with three Ge sites as seen in the crystal structure of UGe$_2$. Note that, due to the asymmetry parameter $\eta\sim 1$ at the Ge1 site (see text), the NQR spectrum for the Ge1 site consists of two main peaks, whereas each spectrum at the Ge2 and Ge3 sites does from four peaks, although the lowest and highest peak for the Ge2 and Ge3 sites are out of the observation range, respectively. The respective NQR spectra (b) and (c), which are obtained at $T=28$ and 4.2 K below $T_{Curie}=31$ K, are affected by internal field $H_{int}$ induced by the onset of ferromagnetism. Dotted lines trace the change in NQR frequencies caused by the increase in $H_{int}$ upon cooling. (d) The $T$ dependence of $1/T_1T$ at 8.50 MHz. The solid curve is a calculation based on the self-consistently renormalized (SCR) spin-fluctuations theory for weakly itinerant ferromagnets.\cite{Moriya}}
\end{figure}

Figure~2 shows the $T$ dependence of $^{73}$Ge-NQR spectrum at $P=1.2$ GPa where the ferromagnetic transition temperature $T_{Curie}$ is decreased to $T_{Curie}=31$ K, which was determined by ac-$\chi$ measurement. In Fig.~2, the NQR spectrum (a) is measured at $T=45$ K for the paramagnetic state, and the respective spectra  of (b) and (c) at $T=28$ and 4.2 K in the ferromagnetic state. The spectrum (a) reveals a structure consisting of well separated peaks associated with three inequivalent Ge sites in one unit cell (see the crystal structure of Fig.~2).
A Ge1 site is closely located along an uranium-zigzag chain. 
The other two Ge2 and Ge3 sites are located out side of this zigzag chain. The number of Ge1 sites is twice as large as the Ge2 and Ge3 sites in one unit cell.
The respective values of the asymmetry parameter $\eta$ of the electric field gradient (EFG) at the Ge1, Ge2 and Ge3 sites were calculated on the basis of band calculation by one of authors (H. Harima) to be $\eta=0.95$ 0.68 and 0.72. Note that a $^{73}$Ge-NQR spectrum with a nuclear spin of $I=9/2$ consists of four  equally separated peaks in case of symmetric EFG, whereas in case of asymmetric EFG, they are no longer equally separated. Especially for $\eta=1$, the four peaks collapse into two peaks. The two large peaks around 6 and 8.5 MHz in the NQR spectrum (a) are assigned to the Ge1 site with the parameters of an NQR frequency $\nu_Q\sim 2.3$ MHz and $\eta=0.98$, consistent with the calculation for the Ge1 site. Note that the peak at 6 MHz (8.5 MHz) is arising from  $\pm 3/2 \iff \pm 5/2$ and $\pm 5/2 \iff \pm 7/2$ transitions ($\pm 1/2 \iff \pm 3/2$ and $\pm 7/2 \iff \pm 9/2$ transitions). Other peaks are reasonably assigned to arise from the Ge2 and Ge3 sites as indicated in the figure, allowing us to deduce $\eta=0.74$ and 0.80, although these values are somewhat larger than the calculated values $\eta=0.68$ and 0.72, respectively. These assignments are also corroborated by the spectrum at $T=4.2$ K and $P=2.0$ GPa as seen in the inset of Fig.~4(a) where FM is completely suppressed.\cite{Harada}
Thus, three Ge sites are separately noticed in the $^{73}$Ge-NQR spectrum observed in the range $f=5-12$ MHz.

Below $T_{Curie}=31$ K at $P=1.2$ GPa, the NQR spectrum undergoes a marked change upon cooling as seen in Figs.~2(b) and 2(c). This is because the onset of FM induces an internal field $H_{int}$ at the Ge sites causing Zeeman-splitting in each of Ge-NQR spectra. Here, we focus on the change in the NQR spectrum under the presence of $H_{int}$ at the Ge1 site which is closely located  to the uranium zig-zag chain. In fact, the dotted lines in Fig.~2 trace the change in NQR frequencies as $H_{int}$ increases at the Ge1 site upon cooling. The NQR spectrum in Fig.~2(a), that consists of two peaks for the Ge1 site in the PM state, splits into multi-NQR lines. From the comparison between experiment and calculation, the respective values of $H_{int}=0.32$ T and $0.88$ T are tentatively estimated at $T=28$ and 4.2 K for the Ge1 site. When noting that the NQR spectra for FM are seemingly observed around the same frequencies as those for the PM state, one may suspect that some non-magnetic sites remain separated spatially in the sample even below $T_{Curie}$. In order to check this possibility on a microscopic level, the nuclear spin-lattice relaxation rate $^{73}(1/T_1)$ was measured at $f= 8.5$ MHz where the NQR spectrum for the Ge1 site has its strongest peak in the PM state. As described in details in the next section,
$T_1$ was determined by the theoretical curve of nuclear magnetization for $\pm 1/2 \iff \pm 3/2$ and $\pm 7/2 \iff \pm 9/2$ transitions where the value of $\eta = 1$ was incorporated.\cite{Chepin}
As indicated in Fig.~2(d), a clear drop in $1/T_1$ below $T_{Curie}$, associated with the suppression of low-lying magnetic fluctuations, revealed that all the Ge sites in the sample are affected by the onset of FM. It is furthermore noteworthy that the $T$ dependence of $1/T_1$ in the PM state obeys a relation of $T/(T-T_{Curie})$ which is the $T$ variation predicted by the self-consistently renormalized (SCR) spin-fluctuations theory for weakly itinerant ferromagnets.\cite{Moriya} 

\begin{figure}[h]
\centering
\includegraphics[width=8.5cm]{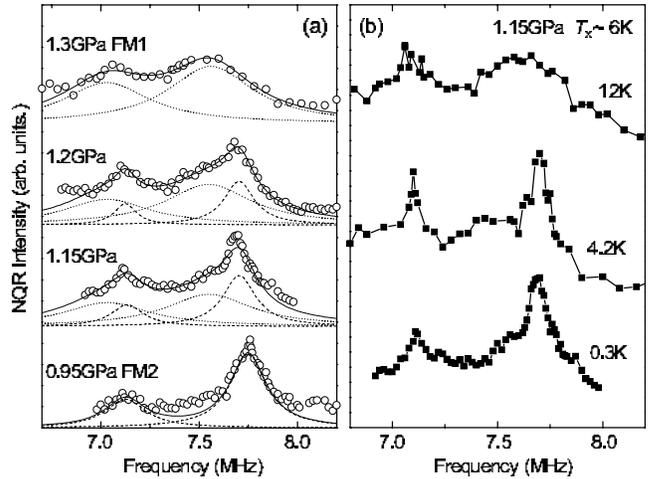}
\caption[]{(a) The $P$ dependence of the $^{73}$Ge spectra at $\sim 0.3$ K. The spectrum for FM1 (FM2) at $P=1.3$ (0.95) GPa is simulated by the overlap of two broad (sharp) Lorentzian spectra as indicated by dotted (dashed) lines. The spectra at $P=1.15$ and 1.2 GPa are reproduced by the superposition of the sharp (dashed) and broad (dotted) two Lorentzian spectra for FM2 and FM1, respectively, demonstrating that phase separation takes place. (b) The $T$ dependence of the spectrum at $P=1.15$ GPa. The sharp peak associated with FM2 appears suddenly below $T_x \sim 6$ K around which the transition from FM1 to FM2 is of first order.}
\end{figure}

Figure~3(a) shows the $P$ dependence of spectrum in the $f = 7-8$ MHz range for the Ge1 site at $T=0.3$ K for $P=0.95$, 1.15, 1.2 and 1.3 GPa. The spectrum for FM1 at $P=1.3$ GPa is significantly broader than that for FM2 at $P=0.95$ GPa. Each spectrum can be accurately modeled  by taking into account the overlap of two Lorentzian spectra. Their full width at a half maximum $\Delta f$ is twice as large for FM1 ($\Delta f \sim 0.53$ MHz) than for FM2 ($\Delta f \sim 0.20$ MHz) as seen in Fig.~3(a). A remarkable finding is that the spectra at $P=1.15$ and 1.2 GPa are reproduced by the superposition of the sharp and broad two Lorentzian spectra for FM2 and FM1, respectively, demonstrating  a phase separation. Here, a fraction of FM2 to FM1 is estimated as ${\rm FM2}:{\rm FM1}\sim 5\pm 1:5\mp 1$ and $2\pm 1:8\mp 1 $ at $P=1.15$ and 1.2 GPa, respectively. This evidences that the $P$-induced transition from FM2 to FM1 is of first order, consistent with other experimental results \cite{Pfleiderer2,Settai}.
When noting an inevitable distribution of pressure $\Delta P\sim 0.03-0.04$ GPa at P = 1.15 and 1.2 GPa, if a critical pressure for first-order transition $P_x$ were in the range 1.15 and 1.2 GPa, it could not be ruled out that this distribution of $\Delta P\sim 0.03-0.04$ GPa makes the phase separated into FM2 and FM1 in association with the first-order transition from FM2 to FM1 at $P_x$.

Figure~3(b) displays the $T$ dependence of the spectrum at $P=1.15$ GPa  where the transition temperature $T_x$ from FM1 to FM2 is estimated as $T_x \sim 6$ K from other measurements \cite{Tateiwa2}. 
The broad spectrum for FM1 at 12 K above $T_x$ resembles that at $P=1.3$ GPa and $T=0.3$ K as seen in the top of Fig.~3(a).
As seen in the spectrum at $T=4.2$ K below $T_x \sim 6$ K, on the other hand, the sharp spectrum appears suddenly,~associated with the first-order transition from FM1 to FM2.
This result, therefore, suggests that the distribution of $H_{int}$ for FM1 is smaller than that for FM2, leading to an increase in magnetization below $T_x$.
It should be noted that even though both FM1 and FM2 are separated around $P_x \sim 1.2$ GPa, SC reveals a highest value of $T_{sc}=0.7$ K. By contrast, when entering FM2 at $P=0.95$ GPa and FM1 at $P=1.3$ GPa apart from $P_x$, $T_{sc}$ goes down for the single phase of either FM1 or FM2. 
\subsection{Nuclear spin-lattice relaxation $-$Evidence for uniform coexistence of ferromagnetism and unconventional superconductivity$-$}

At the PM state, $T_1$ was uniquely determined by a theoretical curve for the recovery of nuclear magnetization $m(t)$ where the respective values of asymmetry parameters was incorporated \cite{Chepin};\\
%
$m_{Ge1,4\nu_Q}(t)=\frac{M(\infty)-M(t)}{M(\infty)}$\\
$=0.075exp\left(-\frac{3t}{T_1}\right)+0.347exp\left(-\frac{7.5t}{T_1}\right)$\\
$+0.425exp\left(-\frac{16.5t}{T_1}\right)+0.153exp\left(-\frac{27.6t}{T_1}\right)\hspace{10mm}$ (1)\\
for the 4$\nu_Q$ transition with $\eta=1$ at the Ge1 site, and \\
$m_{Ge3,1\nu_Q}(t)=\frac{M(\infty)-M(t)}{M(\infty)}$\\
$=0.058exp\left(-\frac{3t}{T_1}\right)+0.357exp\left(-\frac{7.6t}{T_1}\right)$\\
$+0.397exp\left(-\frac{16.3t}{T_1}\right)+0.188exp\left(-\frac{28t}{T_1}\right)$\\
for the 1$\nu_Q$ transition with $\eta=0.8$ at the Ge3 site.
Here ${M(\infty)}$ and ${M(t)}$ are the respective values of nuclear magnetization at the thermal equilibrium state and at time $t$ after  saturation pulses. 

Figure 4(a) indicates $m_{Ge1,4\nu_Q}(t)$ for the 4$\nu_Q$ transition at 8.5 MHz for the Ge1 site that is denoted by an arrow in the spectrum in the inset. The theoretical curve for it is shown by the solid line, which is consistent with a single value of $(1/T_1)_{Ge1}$ in eq.(1). The $T_1$ for the Ge3 site was measured at the 1$\nu_Q$ transition at 10.7 MHz with a single value of $(1/T_1)_{Ge1}$(not shown), being six times larger than for the Ge1 site, i.e. $(1/T_1)_{Ge1}/(1/T_1)_{Ge3} = 6$. 

At the ferromagnetic state, the NQR spectrum splits into multi-NQR lines as seen in Fig.~2(c). $T_1$ was measured at 7.75 MHz where the NQR spectrum is expected to dominantly arise from the Ge1 site. However, $m_{Ge1}(t)$ was not uniquely fitted by eq.(1).  ~Instead, as expected from the change in the spectrum from Fig~.2(a) to Fig.~2(c), by assuming a possible overlap with the spectrum for the Ge3 site, the observed nuclear magnetization $m_{obs}(t)$ is well fitted by \\
$m_{obs}(t)=A\times m_{Ge1,4\nu_Q}+B\times m_{Ge3,2\nu_Q}$\\
$=A\times\left\{0.075exp\left(-\frac{3t}{T_1}\right)+0.347exp\left(-\frac{7.5t}{T_1}\right)\right\}$\\
$+A\times\left\{0.425exp\left(-\frac{16.5t}{T_1}\right)+0.153exp\left(-\frac{27.6t}{T_1}\right)\right\}$\\
$+B\times\left\{0.029exp\left(-\frac{3t}{6T_1}\right)+0.026exp\left(-\frac{7.6t}{6T_1}\right)\right\}$\\
$+B\times\left\{0.157exp\left(-\frac{16.3t}{6T_1}\right)+0.788exp\left(-\frac{28t}{6T_1}\right)\right\}$\\
with $A=0.9$, $B=0.1$ and $(1/T_1)_{Ge1}/(1/T_1)_{Ge3} = 6$ that is indicated by the solid line in Fig.~4(b). Note that the spectrum at 7.75 MHz is also affected by the $2\nu_Q$ transition for the Ge1 site as seen from the change in the spectrum from Fig.~2(a) to Fig.~2(c). By incorporating $m_{Ge1,2\nu_Q}$, the recovery curve indicated by a dashed line is calculated from  the formula of $A_{4\nu_Q}\times m_{Ge1,4\nu_Q}(t)+A_{2\nu_Q}\times m_{Ge1,2\nu_Q}(t)+B\times m_{Ge3,2\nu_Q}(t)$ with $A_{4\nu_Q}+A_{2\nu_Q}=0.9$ where $A_{4\nu_Q}/A_{2\nu_Q}=7/3$ and $B=0.1$, resembling the solid curve. Although the NQR spectrum is seemingly complicated below $T_{Curie}$, the value of $T_1$ at the Ge1 site is reasonably determined and remains reliable.

It should be noted from Fig.~4(c) that the $m_{obs}(t/T_1)$ at $P=1.2$ GPa, that is plotted against a $t$ normalized by the value of $T_1$, is on a nearly same single line as that at $P=1.3$ GPa for FM1.  From this fact, the electronic state at $P = 1.2$ GPa is expected to be compatible to that for FM1 beyond $P_x$, even though the phase separation into FM1 and FM2 is observed at $P=1.2$ GPa. 
\begin{figure}[h]
\centering
\includegraphics[width=6cm]{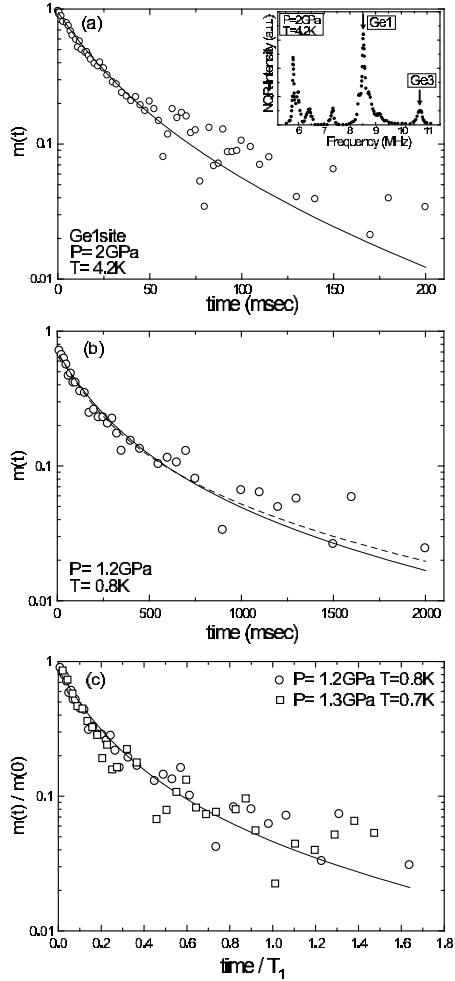}
\caption[]{The observed recovery of nuclear magnetization $m(t)=\frac{M(\infty)-M(t)}{M(\infty)}$. (a) the 4$\nu_Q$ transition (8.5 MHz) at $P=2$ GPa and $T=4.2$ K for the Ge1 site at the paramagnetic phase. The solid line is the theoretical curve with a single value of $T_1$ in eq.(1) (see the text). (b) By assuming a possible overlap with the NQR spectrum for the Ge3 site, the observed nuclear magnetization $m_{obs}(t)$ is well fitted by a solid line which is calculated from the formula of $A\times m_{Ge1,4\nu_Q}(t)+B\times m_{Ge3,2\nu_Q}(t)$ with $A=0.9$ and $B=0.1$, and also by a dashed line by $A_{4\nu_Q}\times m_{Ge1,4\nu_Q}(t)+A_{2\nu_Q}\times m_{Ge1,2\nu_Q}(t)+B\times m_{Ge3,2\nu_Q}(t)$ with $A_{4\nu_Q}+A_{2\nu_Q}=0.9$ where $A_{4\nu_Q}/A_{2\nu_Q}=7/3$ and $B=0.1$. (c) The $m_{obs}(t)$'s at $P= 1.2$ GPa and 1.3 GPa are plotted against a time ($t/T_1$) normalized by the respective values of $T_1$ at $T=0.8$ and 0.7 K.}
\end{figure}

Next we extract characteristics of UGe$_2$ from the thus obtained $T_1$ data. Figures~5 and 6 indicate the $T$ dependence of $1/T_1$ measured at the NQR peak at $f=7.75$ MHz for the spectrum at $P=1.15$ and 1.2 GPa. Note that $1/T_1$ is uniquely determined. Figure~7 measured at the peak for the spectrum for FM1 at $P=1.3$ GPa does that as well.
The $T_1$ measurements probe SC at $T_{sc}= 0.35$, 0.7, and 0.55 K ($\pm 0.05$ K) for $P=1.15$, 1.2, and 1.3 GPa, respectively. In order to confirm the bulk nature of SC for both the phases, $1/T_1T$ at $P=1.2$ GPa was measured in the range $f= 7.75$, 8.5 and 9.12 MHz. 
As indicated in the inset of Fig.~6, all the data reveal a similar $T$ dependence across $T_{sc}$, supporting the homogeneous and bulk nature of SC on a microscopic level.
$1/T_1$ reveals a rapid decrease below $T_{Curie}$, followed by a $T_1T\sim const.$ like behavior upon cooling. In the superconducting state, a clear decrease in $1/T_1$ is evident below $T_{sc}$. Thus, the $T_1$ result, which probes both the transitions into FM and SC, evidences that SC coexists with both FM2 and FM1 on a microscopic level. Markedly, $1/T_1$ decreases without any signature of coherence peak just below $T_{sc}$, which gives evidence for an unconventional nature of SC. 
In fact, the data of $1/T_1$ are well fitted by an unconventional superconducting model with the line-node gap that assumes residual density of states (DOS) $N_{res}$ at the Fermi level as indicated by solid lines in Figs.~5, 6 and 7. A consequence obtained from this model is addressed in next section.
\begin{figure}[ht]
\centering
\includegraphics[width=6cm]{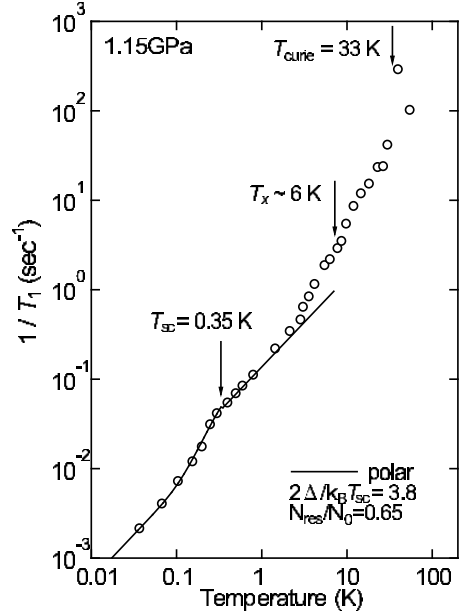}
\caption[]{The $T$ dependence of $1/T_1$ at $P=1.15$ GPa measured at the peak at $f=7.75$ MHz. The solid curve is a calculation based on an unconventional superconducting model with a line-node gap (see the text). The identification of both the phase transitions into SC and FM ensures their uniformly coexistent phase. $T_{Curie}$ was determined by ac-$\chi$ measurement using NMR coil.}
\end{figure}
\begin{figure}[ht]
\centering
\includegraphics[width=6cm]{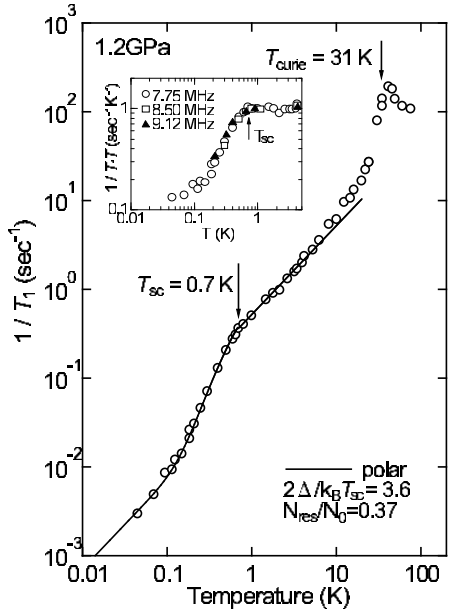}
\caption[]{The $T$ dependence of $1/T_1$ at 1.2 GPa measured at the peak at $f=7.75$ MHz The identification of both the phase transitions into SC and FM ensures their uniformly coexistent phase. The solid curve is a calculation based on an unconventional superconducting model with a line-node gap (see the text). 
The inset shows the frequency dependence of $1/T_1T$ at $P=1.2$ GPa in the range $f=7.75$, 8.5 and 9.12 MHz. The observation of a similar $T$ dependence of $1/T_1T$ ensures the onset of SC over the whole sample.}
\end{figure}
\begin{figure}[ht]
\centering
\includegraphics[width=6cm]{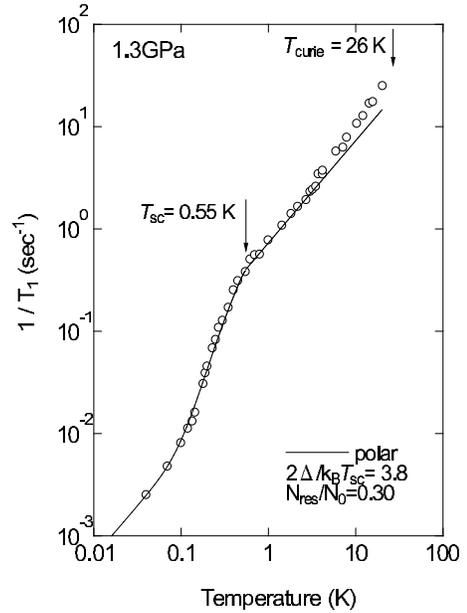}
\caption[]{The $T$ dependence of $1/T_1$ at $P=1.3$ GPa measured at the peak in the spectrum for FM1. The solid curve is a calculation based on an unconventional superconducting model with a line-node gap (see the text).}
\end{figure}
In Fig.~8,  $1/T_1T$ is plotted as function of $T$ at $P=0.95$, 1.15, 1.2 and 1.3 GPa. All the data show a $T_1T=const.$ behavior far below $T_{Curie}$. $1/T_1T$ decreases below $T_{sc}= 0.55$, 0.7, and 0.35 K at $P=1.3$, 1.2, and 1.15 GPa, respectively.
The onset of SC was not confirmed down to $\sim 50$ mK at $P=0.95$ GPa.
A notable point is that the values of $1/T_1T$ above $T_{sc}$ at $P=1.3$ and 1.2 GPa are larger than the values at $P=1.15$ and 0.95 GPa. The underlying electronic state in FM1 seems to possess the larger DOS than in FM2. The value of $(1/T_1T)^{1/2}$ is proportional to the DOS at the Fermi level, in fact, the $P$ dependence of $(1/T_1T)^{1/2}$ scales to that of the $T$-linear coefficient $\gamma$ in the specific heat at low $T$ \cite{Tateiwa2} as indicated in the inset of Fig.~8. 
This underlying electronic state behind the onset of SC is shown to enhance low-lying excitations at the Fermi level near $P_x \sim 1.2$ GPa. As a matter of fact, once $P$ decreases slightly from 1.2 to 1.15 GPa,  $T_{sc}$ decreases dramatically from $T_{sc}\sim 0.7$ K down to 0.35 K. Note that this value of $T_{sc}\sim 0.35$ K is significantly lower than $T_{sc}\sim 0.6-0.7$ K expected from other measurements. From the fact that FM2 makes the NQR peak strong around $f= 7.75$ MHz and its DOS is actually reduced, {\it it is likely that SC for FM2 sets in below $T_{sc}\sim 0.35$ K, whereas SC for FM1 occurs below $T_{sc}\sim 0.6-0.7$ K}.
Therefore, when the phase separation into FM2 and FM1 takes place just below $P_x$, it is expected that the superconducting nature differs at FM2 and FM1 exhibiting SC at $T_{sc}=0.35$ and $0.6-0.7$ K, respectively. In this context, an exact value of $P_x$ may be located in the range  1.15 and 1.2 GPa.
This is because the electronic state at $P=1.2$ GPa is compatible to that for FM1 where the DOS is largely enhanced regardless of the phase separation remaining.
\begin{figure}[h]
\centering
\includegraphics[width=8.5cm]{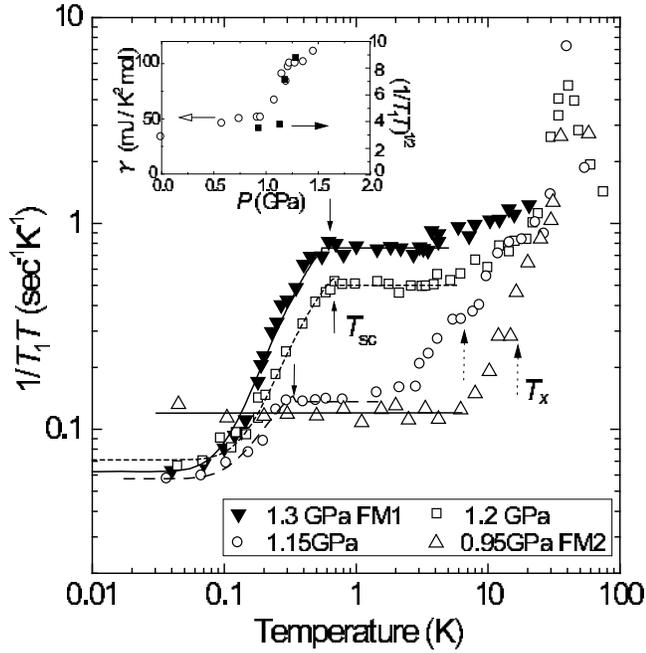}
\caption[]{The $T$ dependence of $1/T_1T$. $1/T_1T$ decreases for FM2 at $P=1.15$ and 0.95 GPa below around $T_x$. The value of $1/T_1T$ for FM1 is significantly larger than those for FM2.  The inset indicates that the $P$ dependence of $(1/T_1T)^{1/2}$ (squares)  scales to that of the $T$-linear electronic contribution in specific heat $\gamma$ (circles). The data of $\gamma$ are referred from ref.\cite{Tateiwa2}}
\end{figure}

\section{Discussion}
As indicated by the solid lines in Figs.~5, 6 and 7 and the dashed lines in Fig.~8, when a line-node gap model is applied with the finite DOS, $N_{res}/N_0$ at the Fermi level, the magnitude of the superconducting energy gap $\Delta$ and $N_{res}/N_0$ are estimated to be $2\Delta/k_BT_{sc}\sim 3.8$, 3.6 and 3.6 with $N_{res}/N_0=0.65$, 0.37 and 0.30 at $P=1.15$, 1.2 and 1.3 GPa, respectively.
~In this model, the origin of $N_{res}/N_0$ cannot be ascribed to some impurity effect because $N_{res}/N_0$ should not depend on $P$.
~In a non-unitary odd-parity pairing model,\cite{Machida,Fomin} a unique relaxation behavior is predicted to depend on the angle between the quantization axis of nuclear-spin system and that of electron-spin one in the non-unitary odd-parity (spin-triplet) SC \cite{Ohmi}. When the former axis is parallel to the latter one, a dependence of $1/T_1\sim T^{2}$ is expected at low $T$, which is inconsistent with the behavior of $1/T_1\sim T^{2.2}$ below $T_{sc}$ at $P_x \sim$1.2 GPa. At the present stage, however, since this issue cannot be resolved experimentally, further analysis on the basis of this model is not yet possible . 

Finally, we wish to remark why SC emerges around the critical point for the first-order transition from FM2 to FM1 around $P_x$. These new phenomena observed in UGe$_2$ should be understood in terms of first-order quantum phase transitions at which the system may fluctuate between states that are separated by a potential barrier. In Fermion systems, if the magnetic critical temperature $T_x$ is suppressed at $P_x$, it involves the diverging magnetic density fluctuations inherent at the critical point from FM2 to FM1 in the quantum Fermi degeneracy region. The Fermi degeneracy by itself generates various instabilities noted as the Fermi surface effects, one of which is a superconducting transition. On the basis of a general argument on quantum criticality, it is shown that the coexistence of the Fermi degeneracy and the critical density fluctuations yield a new type of quantum criticality \cite{Imada}. This makes the physics of first-order quantum phase transitions an extremely rich challenge in both theoretical and experimental studies.

From another context, it is predicted that $T_x$ can be identified with the formation of a simultaneous charge- and spin-density wave (CSDW) and hence near the critical point of this transition, the superconducting pairing is mediated by CSDW fluctuations.\cite{Watanabe} Extensive neutron diffraction studies, however, did not succeed in detecting any static order due to the CSDW and also, the present NQR experiment did not provide possible evidence for the onset of the CSDW below $T_x$. ~The results on UGe$_2$ deserve further theoretical investigations.
\begin{figure}[h]
\centering
\includegraphics[width=8.5cm]{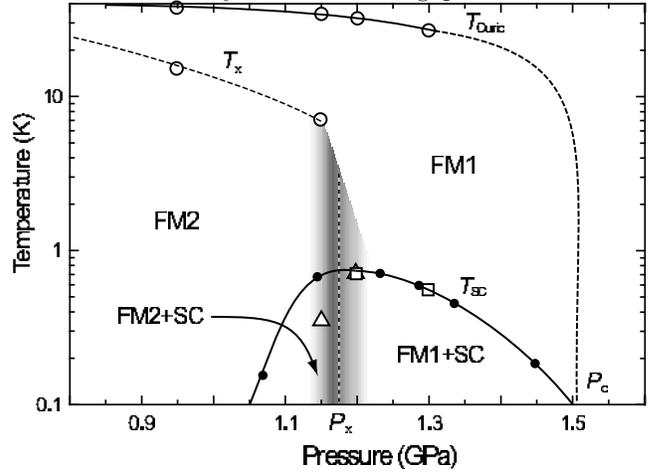}
\caption[]{The $P-T$ phase diagram determined by the present experiment. $T_{sc}$ (solid circle) referred from the previous works is compared with the respective $T_{sc}$ for FM2 and FM1, which are plotted by open triangle and square marks.
At $P=1.2$ GPa, $T_{sc}=0.7$ K coincides for both FM2 and FM1 (see text), whereas at $P=1.15$ GPa just below $P_x$, $T_{sc}=0.35$ K for FM2 is significantly reduced than $T_{sc}\sim 0.6-0.7$ K for resistivity measurement. $P_x$ is tentatively denoted by the dashed line at an intermediate value of 
$P$ between 1.15 and 1.2 GPa. In a shaded region around $P_x$, the phase separation into FM1 and FM2 takes place.  
}
\end{figure}

\section{Summary}
The $^{73}$Ge-NQR measurements in UGe$_2$ have revealed the bulk nature of superconductivity which coexists with the ferromagnetism on the microscopic level. The $P$ dependence of NQR spectrum has unraveled that the $P$-induced magnetic transition is of first order around $P_x$, showing that there is not a quantum critical point around $P_x$.  

The phase diagram determined by the present Ge-NMR measurement is shown in Fig.~9 where the respective $T_{sc}$ for FM2 and FM1 are plotted by the open triangle and square marks. The phases at $P=1.15$ and 1.2 GPa  are separated into FM2 and FM1 in association with an inevitable $P$ distribution $\Delta P\sim 0.03-0.04$ GPa in the cell.  At $P=1.15$ GPa just below $P_x$ where $T_x \sim 6$ K, $T_{sc}=0.35$ K for FM2 is significantly reduced than $T_{sc}\sim 0.6$ K for FM1 in connection with a phase separation into FM2 and FM1. By contrast, SC with the maximum value of $T_{sc}=0.7$ K at $P=1.2$ GPa has been demonstrated to take place for both FM1 and FM2 under the background of the phase separation remaining. As $P$ increases to $P=1.3$ GPa, $T_{sc}$ for FM1 goes down to 0.55 K. The occurrence of SC under the background of FM thus seems to be relevant with the first-order transition at $P_x$.

The $T$ dependence of $1/T_1$ below $T_{sc}$ has been found to be well fitted by the line-node gap model with the $N_{res}/N_0$ at the Fermi level. The large $P$ dependence of $N_{res}/N_0$ cannot be ascribed to  some impurity effect. If the presence of a self-induced vortex state were responsible for the $T_1T=const.$ well below $T_{sc}$, the $P$-induced variation in the DOS at the normal state should cause the $P$ dependence of $T_1T=const.$ below $T_{sc}$. It was not the case. Further experiments are required for understanding novel superconducting characteristics and for addressing a possible order-parameter symmetry in UGe$_2$, either a unitary- or a non-unitary spin-triplet pairing state.

\section{Acknowledgement}
The authors wish to thank K.~Machida, N.~Tateiwa, T. C. Kobayashi, G. -q. Zheng, A.~D.~Huxley and J. Flouquet for helpful discussions.
This work was supported by Grant-in-Aid for Creative Scientific Research（15GS0213), MEXT and the 21st Century COE Program supported by Japan Society for the Promotion of Science.  H.K. and S.K.have been supported by Research Fellowship of the Japan Society for the Promotion of Science for Young Scientists.

\end{document}